\documentclass[prb,aps,twocolumn,superscriptaddress,showpacs]{revtex4}

\usepackage{graphicx}

\begin{document}

\title{Universal Static and Dynamic Properties of the Structural 
Transition in Pb(Zn$_{1/3}$Nb$_{2/3}$)O$_{3}$}

\author{C. Stock}

\affiliation{Department of Physics, University of Toronto, Ontario, Canada 
M5S 1A7}

\author{R. J. Birgeneau}

\affiliation{Department of Physics, University of Toronto, Ontario, Canada 
M5S 1A7}

\author{S. Wakimoto}

\affiliation{Department of Physics, University of Toronto, Ontario, Canada 
M5S 1A7}

\author{J. S. Gardner}

\affiliation{National Research Council, Chalk River, Ontario, Canada, K0J 
1J0}

\author{W. Chen}

\affiliation{Department of Chemistry, Simon Fraser University, Burnaby, 
British Columbia, Canada V5A 1S6}

\author{Z. -G. Ye}

\affiliation{Department of Chemistry, Simon Fraser University, Burnaby, 
British Columbia, Canada V5A 1S6}

\author{G. Shirane}

\affiliation{Physics Department, Brookhaven National Laboratory, Upton, 
New York 11973}

\date{\today}

\begin{abstract}

The relaxors Pb(Zn$_{1/3}$Nb$_{2/3}$)O$_{3}$ (PZN) and 
Pb(Mg$_{1/3}$Nb$_{2/3}$)O$_{3}$ (PMN) have very similar properties based 
on the dielectric response around the critical temperature $T_{c}$ 
(defined by the structural transition under the application of an electric 
field).  It has been widely believed that these materials are quite 
different below $T_{c}$ with the unit cell of PMN remaining cubic while in 
PZN the low temperature unit cell is rhombohedral in shape.  However,  
this has been clarified by recent high-energy x-ray studies which have shown 
that PZN is rhombohedral only 
in the skin while the shape of the unit cell in the bulk is nearly cubic.  
In this study we have performed both neutron elastic and inelastic scattering to show 
that the temperature dependence of both the diffuse and phonon scattering 
in PZN and PMN is very similar.  Both compounds show a nearly 
identical recovery of the soft optic mode and a broadening of the acoustic 
mode below $T_{c}$.  The diffuse scattering in PZN is suggestive of an 
onset at the high temperature Burns temperature similar to that in PMN.  
In contrast to PMN, we observe a broadening of the Bragg peaks in both the 
longitudinal and transverse directions below $T_{c}$.  We reconcile 
this additional broadening, not observed in PMN, in terms of structural 
inhomogeneity in PZN.  Based on the strong similarities between PMN and PZN, we 
suggest that both materials belong to the same universality class and discuss the 
relaxor transition in terms of the three-dimensional Heisenberg model with cubic anisotropy 
in a random field.  

\end{abstract}


\maketitle

\section{Introduction}

\begin{figure}[t]

\includegraphics[width=8cm] {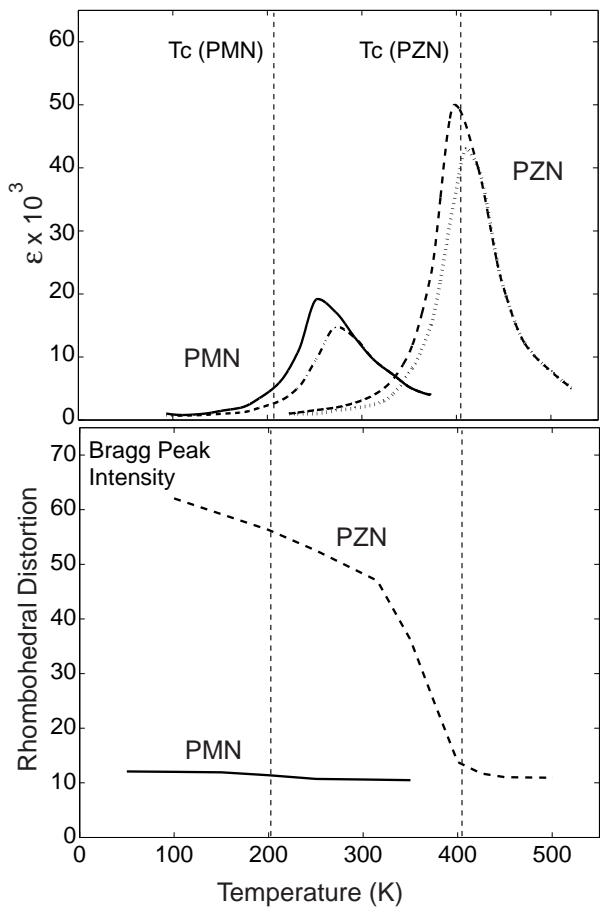}

\caption{\label{figure1} Schematic figure showing the temperature 
dependence of the dielectric constant (data taken from Ye \textit{et al.}) 
and the Bragg peak intensity for both PMN (220) and PZN (300) which 
characterize rhombohedral distortion.  The upper 
figure shows the dielectric constant for 100 Hz (peaked at lower 
temperatures) and 1 MHz (peaked at higher temperatures). The dotted lines 
indicate the critical temperature measured under the application of an 
electric field.}

\end{figure}

	The relaxor materials of the chemical form PbBO$_{3}$ have generated much 
interest recently due to their large piezoelectric constants which are an 
order of magnitude larger than those of conventional 
ferroelectrics.~\cite{Park97:82, Ye98:155}  These materials are 
characterized by quenched disorder on the B site and display a diffuse 
phase transition with a broad and frequency dependent dielectric response 
peaked at the temperature $T=T_{max}$ (reproduced in the upper panel of 
Fig. \ref{figure1}).~\cite{Kuwata79:22} Early studies of the refractive 
index showed that regions of local ferroelectric order are formed in a 
paraelectric background at a temperature $T=T_{d}$ (denoted as the Burns 
temperature).~\cite{Burns83:10}  The nature of this diffuse transition and 
the low temperature ground state has been the focus of many recent 
studies. Most recently, much attention has centered on the material 
Pb(Mg$_{1/3}$Nb$_{2/3}$)O$_{3}$ (PMN) with a relatively accessible Burns 
temperature of $T_{d}\sim$ 620 K.

	High temperature neutron inelastic scattering was originally 
investigated 
by Naberezhnov \textit{et al.}~\cite{Naberezhnov99:11} and was followed by 
Gehring \textit{et al.}~\cite{Gehring01:87} who clearly showed the 
presence of a soft ferroelectric mode in PMN which becomes overdamped 
below $T_{d}$.~\cite{Gehring01:63}  Diffuse scattering studied by 
Hirota \textit{et al.}~\cite{Hirota02:65} showed that the growth of diffuse 
scattering could be well described by the formation of {\it phase-shifted} 
polar nanoregions in which the atomic displacement was described by two 
components - a center of mass conserving component and another 
non-conserving component associated with a uniform phase shift.  The connection 
between the diffuse and phonon scattering was later established through a 
detailed phonon study~\cite{Wakimoto02:xx} which showed the soft optic and 
the transverse acoustic modes to be strongly coupled. This coupling 
provides a natural explanation for the presence of phase-shifted polar 
nanoregions as the center of mass conserving atomic shift can be 
attributed to the condensation of the soft optic mode while the 
non-conserving component corresponds to a softening of the acoustic mode.  

	Unpoled PMN shows essentially no structural change through 
$T_{c} = 213$~K with only a small transverse broadening in the Bragg 
peaks.~\cite{Wakimoto02:xx}  The transition temperature $T_{c}$ in PMN is 
defined by the structural transition after poling the sample under an 
electric field.~\cite{Ye93:83}  Despite no structural phase transition the soft 
optic mode in unpoled PMN was found to recover below $T_c$ and 
the energy squared was observed to 
increase linearly with decreasing temperature.~\cite{Wakimoto02:65} 
This is 
extremely surprising as such a recovery has previously been thought to be 
associated with a well defined structural transition.~\cite{Cochran69:18}  
The recovery of the soft mode suggests the presence of a ferroelectric 
distortion in PMN, 
despite the absence of any change in the average unit cell shape. 

	Until very recently, Pb(Zn$_{1/3}$Nb$_{2/3}$)O$_{3}$ (PZN) 
and PMN were thought to be very different with 
PMN remaining cubic at all temperatures and PZN undergoing a structural 
phase transition to a rhombohedral phase~\cite{Nomura69:27} at a very well 
defined critical temperature $T_{c} = 410$~K.  The Bragg scattering around $T_{c}$ 
in both these compounds displays contrasting behavior as illustrated in 
the lower panel of Fig.~\ref{figure1} which shows a large increase in the 
neutron Bragg peak intensity for PZN at $T_{c}$ but no change in PMN over 
a broad range in temperature.  The structural transition in PZN has been 
reexamined in detail by a recent diffraction study using 
$\sim 9$~keV x-rays~\cite{Lebon02:14}.  Studies of the diffuse scattering in PZN 
by La-Orauttapong \textit{et al.} showed that the diffuse scattering 
started near $T_{c}$, well below $T_{d}$ where the onset 
of diffuse scattering occurs in PMN, highlighting another difference 
between PMN and PZN.~\cite{DLa01:64} 

	The notion that the unit cell in PZN undergoes a transition from a cubic 
to rhombohedral in shape at a well defined transition temperature $T_{c}$ 
changed with a recent high-energy and penetrating x-ray study by Xu 
\textit{et al.}~\cite{Xu02:xx} and with a neutron diffraction study on 
PbTiO$_3$ (PT) 
doped PZN by Ohwada \textit{et al.}~\cite{Ohwada02:xx}.  Both these 
studies found that the low temperature structural phase was not 
rhombohedral but rather was characterized by a cubic unit cell, denoted as phase X.  
The 67 keV high energy x-ray study revealed phase X to be characterized by 
resolution limited Bragg peaks suggesting a well ordered 
low-temperature structure.  The connection to the PMN system was made 
recently by a high resolution neutron diffraction study on 10 $\%$ PT 
doped PMN~\cite{Gehring289} which found the low temperature ground state 
to also be characterized by a cubic shaped unit cell with nearly 
resolution limited Bragg peaks.  These recent studies on PZN and PT doped 
PMN show that PZN and PMN may not be as different as once thought.  These 
results necessitate a more detailed study of the dynamical and static 
properties of pure PZN and a comparison between PZN and PMN. 

	To investigate the analogy between PMN and PZN we have conducted a 
detailed study of the structural and dynamical properties of PZN around the 
critical temperature $T_{c}$ using neutron inelastic and elastic 
scattering.  We will define the similarities between PMN and PZN through 
both the inelastic scattering from the phonons and the elastic diffuse 
scattering.  A key difference between these relaxor systems illustrated by 
the Bragg scattering will be discussed and reconciled in terms of 
structural inhomogeneity in PZN.  The strong similarity between these two materials 
suggests they belong to the same universality class.  We propose that the relaxor 
transitions can be understood in terms of the three-dimensional Heisenberg model with cubic anisotropy  
in the presence of a random field.  This model is very attractive as it 
unifies the many temperature scales measured by various experimental probes.

\section{Experimental Details}

	Neutron scattering experiments were performed at the NRU reactor, Chalk 
River Laboratories.  Elastic scattering studies of the diffuse scattering and the 
Bragg peaks were conducted on the N5 spectrometer by holding the incident 
and final energies fixed at 14.8 meV.  Diffuse scattering results were obtained using 
a flat Graphite (002) monochromator and analyzer.  Horizontal collimation 
was set by Soller slits from the reactor to detector at 
77$'$-33$'$-\textit{S}-35$'$-60$'$ (where \textit{S} denotes the sample) and a 
Graphite filter was used on the scattered side to remove higher order 
neutrons.  For elastic studies of the Bragg peaks we have matched the 
$d$-spacings of the sample to that of the (220) reflection of a Germanium 
analyzer with a small mosaic of $\sim$3$'$.  By matching the $d$-spacings of 
the sample and analyzer crystal we have obtained good resolution allowing 
us to study the small broadening of the Bragg peaks below $T_{c}$.  For elastic 
measurements of the Bragg peaks we have also set the horizontal 
collimation to 77$'$-6$'$-\textit{S}-12$'$-60$'$.

	Inelastic measurements studying both the transverse acoustic and optic 
modes were conducted on the C5 spectrometer by fixing the final energy to 
14.5 meV and varying the incident energy.  A variable focusing graphite 
(002) monochromator and a flat graphite analyzer were used with the 
horizontal collimation set at 33$'$-29$'$-\textit{S}-48$'$-72$'$ to study 
the acoustic mode and 33$'$-48$'$-\textit{S}-51$'$-72$'$ to study the 
optic mode.  A graphite filter was also used on the scattered side to 
filter out higher order neutrons and a cold sapphire filter was used before 
the monochromator to reduce the fast neutron background.

	Single crystals of PZN were grown by spontaneous nucleation from high 
temperature solution using PbO as flux, based on the technique previously 
described.~\cite{Zhang00:78} A single crystal with a volume of 0.2 cc and 
a weight of 1.8 g was selected for the experiments. It was oriented with 
the largest face parallel to the (100) cubic plane. The sample of PZN was 
mounted, using tantalum wire, to a boron-nitride post with a copper insert 
to ensure good thermal contact.  The sample was mounted in an orange 
cryofurnace such that reflections of the form (HK0) lay in the scattering 
plane.  The room temperature lattice constant was measured to be 4.04 \AA~ 
and therefore 1 reciprocal lattice unit (r.l.u.) corresponds to 
$2 \pi/a$ = 1.55 \AA$^{-1}$.  Because PZN decomposes at very high 
temperatures we have kept our high temperature measurements below 550 K to 
avoid any possible damage.  Unfortunately, this means that the studies 
above or around the Burns temperature $T_{d}$, which exceeds 670 K, are 
not possible in PZN.

\section{Bragg Scattering}

\begin{figure}[t]

\includegraphics[width=8cm] {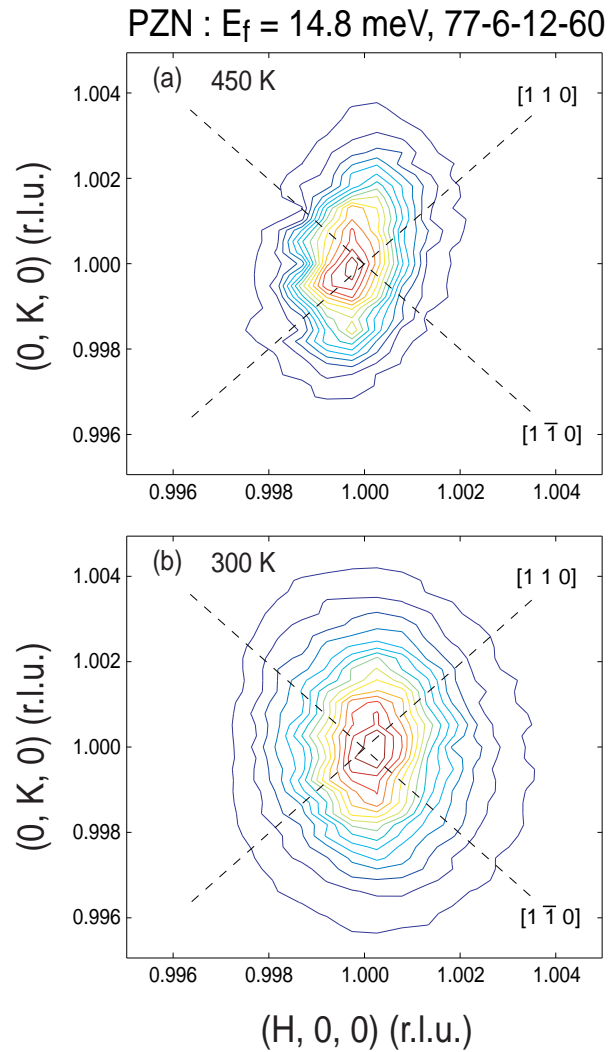}

\caption{\label{bragg_1} Contour plots of the (110) reflection as scanned 
in the \textit{(HK0)} plane above (\textit{a}) and below (\textit{b}) the 
critical temperature $T_{c}$.  These data were taken using a Germanium 
analyzer with a small mosaic spread.  Dashed lines indicate longitudinal
[110] and transverse [1$\overline{1}$0] directions.}

\end{figure}

\begin{figure}[t]

\includegraphics[width=8cm] {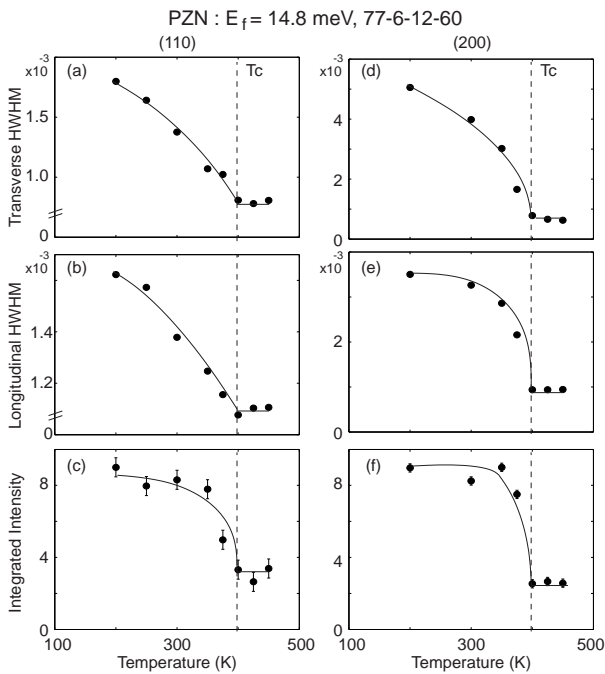}

\caption{\label{bragg_2} Plots of the transverse, longitudinal widths, and 
the integrated intensities as a function of temperature.  The widths were 
obtained from Lorentzian fits to the observed scattering.  In all cases 
there is clear anomaly at the critical temperature $T_{c}$.  The solid 
lines are guides to the eyes.}

\end{figure}

	To investigate the Bragg peak profile as a function of temperature around 
$T_{c}$ we have used a Germanium analyzer with a 
small mosaic spread of $\sim 3'$.  Based on low-energy x-ray results 
showing a transition from cubic to rhombohedral we conducted scans around 
the (110) position.  If a rhombohedral phase exists at low temperatures 
the Bragg peak should split along the longitudinal [110] direction only.  
Instead of such a distortion we observe a broadening of the (110) Bragg 
peak along both the longitudinal [110] and transverse [1$\overline{1}$0] 
directions (Fig.~\ref{bragg_1}). 
This is in contrast to the behaviour in PMN which shows only transverse broadening 
below $T_c$ ~(Ref.~\onlinecite{Wakimoto02:xx}).
We have also investigated the temperature 
dependence of the (200) Bragg peak which is found to also display both 
transverse and longitudinal broadening below $T_{c}$ 
(Fig.~\ref{bragg_2}).  
As mentioned in the next section, we find no observable 
temperature dependent diffuse scattering around the (200) position.
This means that the broadening of the Bragg peaks shown in Fig.~\ref{bragg_2}
cannot be associated with the diffuse component.  

	These results contrast with the results of both low-energy and 
high energy x-ray 
measurements. The $\sim$ 9 keV x-ray study by Lebon \textit{et al.} showed 
the ground state of PZN at low temperatures to have a rhombohedral unit cell.  The 
high-energy 67 keV x-ray results by Xu \textit{et al.} found the unit cell 
to be nearly cubic and to be very well ordered characterized by resolution 
limited Bragg peaks.  Xu \textit{et al.} also confirmed the result of 
Lebon \textit{et al.} by showing that a splitting of the (111) Bragg peak 
characteristic of a rhombohedral distortion existed when a lower incident beam energy of 32 keV was used.  The contrasting results presented by x-ray scattering with various 
incident energies can be reconciled based on the relative penetration depths.  As noted 
by Xu \textit{et al.}, the lower energy x-rays in the range 10-32 keV have 
a penetration depth in PZN of only $\sim$ 10-60 $\mu m$ whereas x-rays with an incident 
energy of 67 keV have a penetration depth of $\sim$ 400 $\mu m$ which 
should give a more accurate picture of the bulk properties.  Therefore, in 
PZN rhombohedral order is formed in the skin of the sample whereas the bulk 
has a well ordered pseudocubic structure.  This new bulk ground state 
has been referred to as Phase-X.

	Our neutron results showing both transverse and longitudinal broadening 
of the Bragg peaks are consistent with the picture obtained from x-rays.  
Since neutrons probe the entire sample volume the observed Bragg peak 
profile is a combination of both the bulk and the near-surface region.   
The presence 
of both transverse and longitudinal broadening is consistent with finite 
size effects resulting from domains.
Thus, based on the x-ray 
results, the broadening along both directions is very likely the result 
of structural 
inhomogeneity in the bulk sample.  The size of these 
domains can be estimated from the half-width (estimated from a Lorentzian 
profile)~\cite{halfwidth} of the Bragg peaks at low temperatures to be 
$\sim$ 200-400 \AA.  This estimate is consistent with the value of 700 \AA\ 
estimated by Xu \textit{et al.} based on the low energy x-ray results.  

	The existence of finite size effects resulting from structural 
inhomogeneity is also consistent with the temperature 
dependence of the integrated intensity displayed in Fig.~\ref{bragg_2} for 
both the (110) and (200) Bragg peaks.  Both Bragg peaks show a large 
increase in the integrated intensity at $T_{c}$ characteristic of a 
release of extinction at low temperatures.  Such an effect is the result 
of a broadening of the overall mosaic spread of the crystal allowing 
better penetration of the neutron beam into and out of the crystal.  
Detailed analysis of extinction and its dependence on the domain size is 
discussed elsewhere.~\cite{Sakata78:34}  

	Thus, our results are consistent with the high-energy x-ray results 
showing the absence of rhombohedral order at low temperatures.  We do 
observe a longitudinal line broadening indicating structural 
inhomogeneity.  In the sense that no rhombohedral order is formed in the 
bulk at low temperatures, the structural properties of PZN are similar to those of 
PMN.

\section{Diffuse Scattering}

\begin{figure}[t]

\includegraphics[width=8cm] {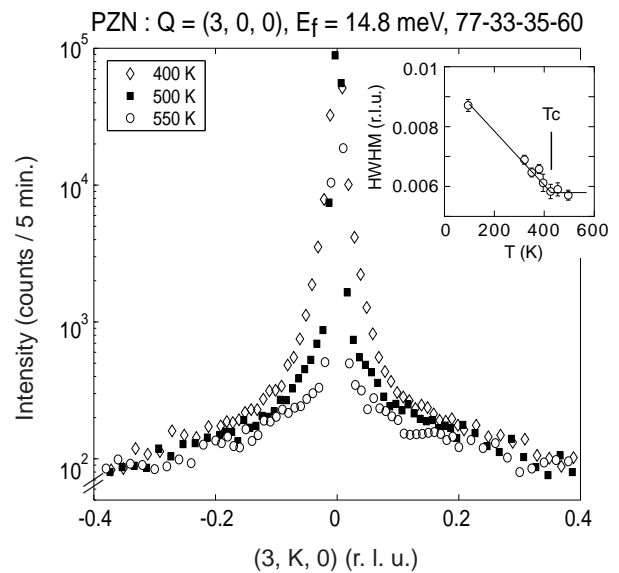}

\caption{\label{diffuse_1}  Diffuse scattering around the (300) position 
at several temperatures.  The diffuse scattering is clearly present well 
above the critical temperature $T_{c} = 410$~K and suggestive that diffuse 
scattering starts at the Burns temperature $T_{d}$.  The inset shows the 
unconvoluted HWHM of the (300) Bragg peak as a function of temperature.}

\end{figure}

	The absence of a clear rhombohedral distortion at low temperatures 
necessitates a revaluation of the diffuse scattering.  Previous diffuse 
scattering results by D. La-Orauttapong \textit{et al.}~\cite{DLa01:64} 
have indicated that the temperature dependent diffuse scattering has an onset 
near $T_{c} \sim 410$~K.  This result contrasts with that in PMN where 
the diffuse scattering starts at the much higher temperature of $T_{d}$ where 
polar nanoregions are formed. 

\begin{figure}[t]

\includegraphics[width=8.0cm] {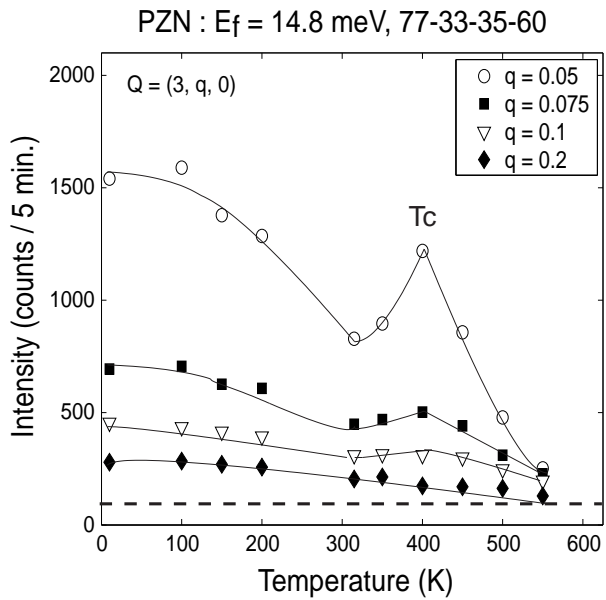}

\caption{\label{diffuse_2} Measured intensity at (3, q, 0) as a function 
of temperature for various values of q.  Large values of q show a gradual 
and continuous growth with decreasing temperature of the diffuse scattering through $T_{c}$ while smaller 
values of q show evidence of critical scattering. The horizontal dotted 
line indicates the constant background.}

\end{figure}

	We have studied the diffuse scattering and its temperature dependence 
around the (300) reflection where the diffuse scattering is strong.  We 
have found that the lineshape of the diffuse scattering at low temperatures 
is qualitatively similar to that of PMN studied in detail by You 
\textit{et al.}~\cite{You97:79} These authors have found that the diffuse intensity in PMN is elongated along both the [110] 
and [1$\overline{1}$0] directions at the (300) position.  
Figure~\ref{diffuse_1} shows the temperature variation of the (300) diffuse 
profile indicating the presence of diffuse scattering even above 500~K.
The presence of a diffuse component above $T_c$ is clearly demonstrated  
in Fig.~\ref{diffuse_2} which shows the temperature dependence of the diffuse 
intensity at different $q$-positions with a constant background
indicated by a dashed line.
It is again clear that diffuse 
scattering is still present at 550 K.  These results indicate that diffuse scattering starts well above 
$T_{c}$ and is suggestive of an onset at the Burns temperature 
$T_{d}$ similar to the case in PMN.
As also displayed in Fig.~\ref{diffuse_1}, this continuous 
growth of the diffuse scattering is in contrast to the transverse HWHM of 
the Bragg peak which shows a sharp anomaly at $T_c$. 
The gradual growth of the diffuse intensity starting at temperatures in excess of 500 K differs with the suggestion 
that the diffuse scattering starts at $\sim$ 450 K~\cite{DLa01:64} and is 
consistent with previous diffuse work on PMN which clearly showed the 
onset of a diffuse component at the Burns 
temperature.~\cite{Hirota02:65,Naberezhnov99:11}

Another important characteristic of diffuse scattering in PZN is shown 
in Fig.~\ref{diffuse_2}.
For large values of \textit{q} the diffuse scattering 
grows continuously through $T_{c}$ but at smaller values of \textit{q} a 
clear peak is observed in the measured intensity suggesting the presence 
of critical scattering.  Even though this is consistent with the diffuse 
scattering in PMN-PT~\cite{Koo02:65} these results differ from the current 
understanding of diffuse scattering in pure PMN which shows a continuous 
change through $T_{c}$.~\cite{Dkhil01:65}  Therefore the diffuse 
scattering around $T_{c}$ could be interpreted as consisting of scattering 
from the polar nano-regions formed at $T_{d}$ and critical scattering from 
some component of the sample undergoing a structural phase transition at 
$T_{c}$ (see Fig~\ref{figure1}).  Based on our analysis of the Bragg 
scattering this extra component of the sample which undergoes a structural 
transition is located in the near-surface region and is the same component studied by 
Lebon \textit{et al.} using low energy x-rays.

	We believe that the structural inhomogeneity is the underlying 
reason for the 
apparent discrepancy between our diffuse scattering results and those of a 
previous diffuse study on pure unpoled PZN by D. 
La-Orauttapong.~\cite{DLa01:64}  In that study, a two phase critical 
scattering model was applied to the diffuse intensity around the (110) 
position which used a Lorentzian function to describe the diffuse 
scattering and a Gaussian to describe the Bragg peak.  This analysis 
showed the onset of the diffuse Lorentzian component near $T_{c}$ and no 
change in the Bragg peak width.  We believe that this model is oversimplified 
as the scattering should be the sum of the diffuse component, the Bragg 
peak, and the critical scattering coming from the skin of the sample which 
does enter a rhombohedral phase at low temperatures.  Therefore we 
speculate that the diffuse Lorentzian component in the fits by D. 
La-Orauttapong \textit{et al.} reflected mostly the critical component 
not characteristic of the bulk phase (phase-X) of the sample.

	To test further the similarity with PMN we have checked for diffuse 
scattering at the (200) and (110) positions at temperatures above and 
below $T_{c}$. We have found no observable change in the scattered 
intensity around the (200) Bragg peak position and only a small change 
around the (110) Bragg peak.  This implies the absence of diffuse 
scattering around (200) and only a weak diffuse component around (110). 
These results for the structure factors are predicted by the phase shifted 
polar nanoregion model used to describe the diffuse scattering in PMN.  In 
this model the atomic displacements are divided into a component which 
conserves the center of mass and another, associated with an overall phase 
shift, which does not.  The component associated with the center of mass conserving optic modes is determined by a linear combination of the Last and Slater modes.~\cite{Harada70:26} Based on a calculation similar to that conducted by Hirota \textit{et 
al.} and assuming the same parameters for the product of the atomic 
displacement and the atomic mass at each lattice site as in PMN we find 
the calculated phase shift for PZN to be $\delta_{shift} \sim 0.56$ 
(normalized to the total shift of the Pb atom). 
This is quite similar to that of PMN, $\delta_{shift} \sim 0.58$.  Hirota {\it et al.} found by choosing the correct weight of both Last 
and Slater modes and using the calculated phase shift, the intensity of the 
diffuse scattering at the (200) position could be made small compared to 
the diffuse scattering at the (110) and (300) positions.  This is the same 
trend observed in PZN.

	The absence of any observable temperature dependent diffuse component 
around the (200) position is also important as it illustrates that the line 
broadening observed in the Bragg scattering is coming from structural 
effects with an onset at $T_{c}$ and not from a diffuse component (see 
Fig.~\ref{bragg_2}).  This result contrasts with the previous suggestion 
that the apparent change in the Bragg peak widths were a result of a 
strong diffuse temperature dependence around $T_{c}$.~\cite{DLa01:64}  
Clearly, the Bragg peaks show an anomalous linewidth broadening beginning at the  
critical temperature $T_{c}$ in PZN (see Fig.~\ref{bragg_2}).

	Our analysis based on the temperature dependence and structure factors 
link the diffuse scattering directly with the formation of polar 
nanoregions.  Despite the fact that we are not able to reach the Burns 
temperature $T_{d} \sim 700$~K our results suggest that the onset of the 
diffuse scattering starts at $T_{d}$ (where polar nanoregions are 
formed).  Our results are strongly analogous to those in PMN where neutron 
scattering has shown the presence of a diffuse component at the accessible 
Burns temperature $T_{d}$ $\sim$ 620 K. 

\section{Lattice Dynamics}

In the previous sections, we have shown that PZN has similar diffuse 
scattering characteristics to those of PMN.  
Another important feature closely related to the relaxor mechanism are 
the properties of the transverse optic and acoustic phonon modes.
These phonon modes in PMN have been 
well studied.  Striking features in PMN appear around $T_c$, such as
recovery of the soft optic mode below $T_c$ and disappearance of 
the acoustic phonon line broadening.
In this section, we report the transverse optic and acoustic phonons in PZN 
measured around $T_c$ and compare in detail these results to those of PMN to illustrate the similarity between these two relaxor systems. 
Firstly, we introduce the lineshape used to analyze our phonon results.

\subsection{Lineshape: Simple Harmonic Oscillator}

\begin{figure}[t]
\includegraphics[width=8cm] {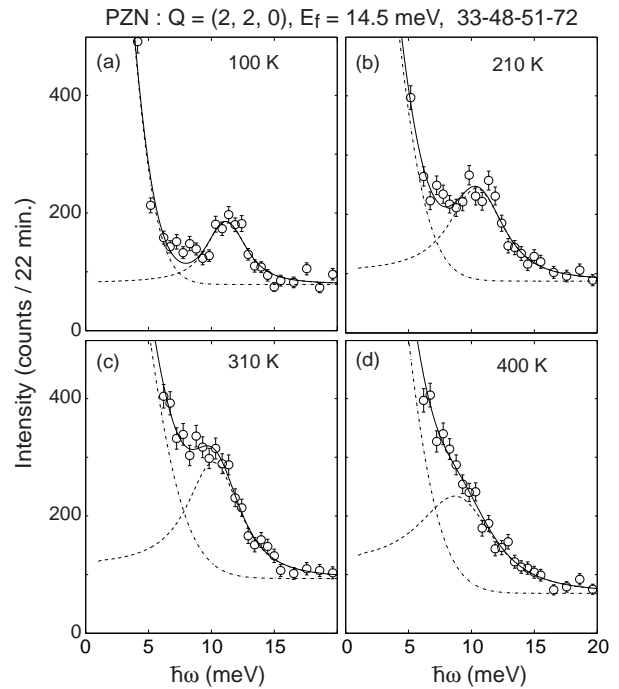}
\caption{\label{optic_1}Scans of the transverse optic mode at (220).  The 
solid line is the sum of the convoluted Lorentzian and a Gaussian centered 
at $\omega$=0.  The dashed line is the Lorentzian contribution and the 
dotted-dashed line is the Gaussian at zero energy transfer.}
\end{figure}

\begin{figure}[t]
\includegraphics[width=8cm] {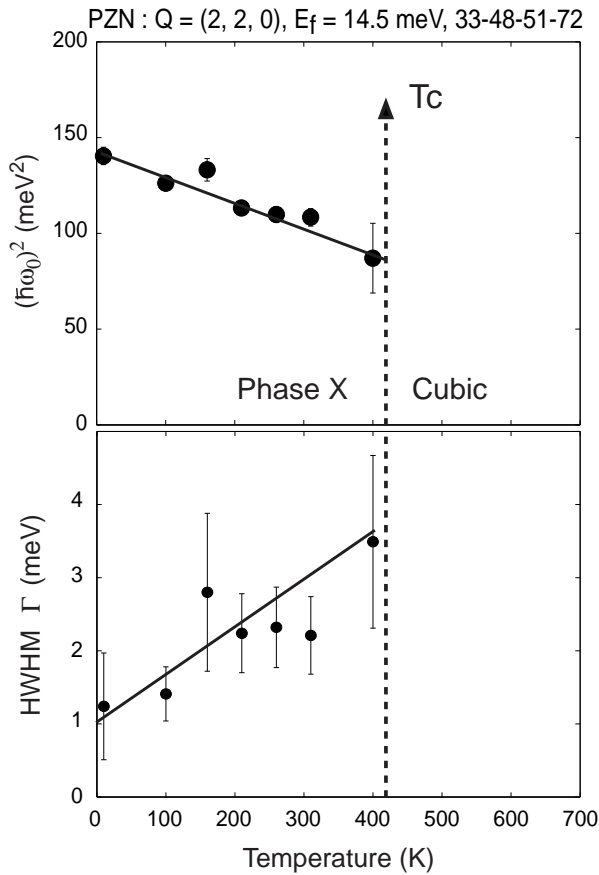}
\caption{\label{optic_2} Transverse optic fitted parameters as a function 
of temperature.  The upper panel is the phonon frequency 
$\omega_{\circ}^{2}$ plotted as a function of temperature.  The HWHM 
$\Gamma$ is plotted in the lower panel.}
\end{figure}

\begin{figure}[t]
\includegraphics[width=8cm] {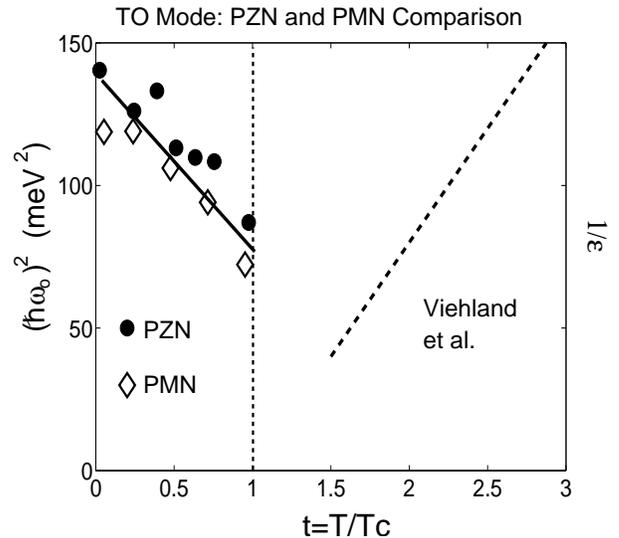}
\caption{\label{optic_3} The soft mode position $(\hbar 
\omega_{\circ})^{2}$ as a function for reduced temperature $t=T/T_{c}$ for 
both PMN and PZN.  The data for both materials overlaps extremely well 
illustrating the strong similarity between these two materials.  The dashed line above \textit{t}=1 is a schematic representation of $1/\epsilon$ taken from Viehland \textit{et al.}}
\end{figure}

	Measurement of the neutron scattering intensity provides a direct measure 
of S({\bf{Q}}, $\omega$), which is related to the 
imaginary part of the susceptibility $\chi$$'$$'$({\bf{Q}}, $\omega$) by 
the fluctuation dissipation theorem~\cite{Shirane:book},

\begin{equation}
S({\bf q},\omega)=\pi^{-1} \left[n(\omega)+1 \right] \chi''({\bf q},\omega),
\end{equation}

\noindent where the energy transferred to the sample is defined by $\hbar 
\omega=E_{i}-E_{f}$ and $n(\omega)=1/(e^{\hbar \omega/k_{B}T}-1)$ is the 
Bose factor. In order to obtain detailed information on the linewidth from 
the inelastic spectrum, a model must be convolved with the resolution 
function and then fit to the observed scattering.  To analyze the phonon 
scattering as a function of temperature we have taken $\chi$$'$$'$ to be 
described by the formula for the damped simple harmonic oscillator.  We 
have used the equation given by the antisymmetrized linear combination of 
two Lorentzians,

\begin{eqnarray}
\label{SHO} \chi''({\bf{q},}\omega)={A\over {[\Gamma(\omega)^{2}+ 
\left(\hbar \omega- \hbar \omega_{\circ}(\bf{q})\right)^{2}}]} \\
- {A\over {[\Gamma(\omega)^{2}+ \left(\hbar \omega+\hbar 
\omega_{\circ}(\bf{q})\right)^{2}}]},
\nonumber
\end{eqnarray}

\noindent where $\Gamma(\omega)$ is the frequency dependent half-width-at-half-maximum (HWHM), $\omega_{\circ}({\bf{q}})$ is the undamped phonon 
frequency, and \textit{A} is the amplitude.  For acoustic modes, we have 
approximated the dispersion to be linear, $\omega_{\circ}({\bf{q}})=c 
|\bf{q}|$, where $c$ is the phonon velocity.  For optic modes, we have set 
the dispersion to have the form 
$\omega_{\circ}({\bf{q}})^{2}=\Omega_{\circ}^{2}+(\Lambda |{\bf{q}}|)^2$, 
where $\Omega_{\circ}$ is the soft-mode energy at $q=0$ and $\Lambda$ is a 
constant for a given $\bf{q}$ direction.~\cite{Shirane70:2,Axe70:1}

	The antisymmetrized Lorentzian lineshape obeys detailed 
balance~\cite{Shirane:book} which requires $\chi$$'$$'$ to be an odd 
function in energy.  By making several substitutions~\cite{Cowley73:6} the 
lineshape in our analysis can be shown to be equivalent to those employed in previous studies 
which used a frequency independent damping constant, 
$\Gamma_{\circ}$.~\cite{Shirane70:2}  The measured intensity was fitted to 
Eq.~\ref{SHO} convolved with the instrumental resolution and a 
non-convoluted Gaussian function was used to describe the component at 
$\omega$=0.  A constant was used to describe the overall background.

\subsection{Soft Optic Mode}

	The soft transverse optic mode was studied by scans with fixed 
${\bf{Q}}$ 
at the (2, 2, 0) position.  Figure \ref{optic_1} displays scans over a 
broad range of temperature below $T_{c}$.  The solid lines are the results of fits to 
the antisymmetrized Lorentzian lineshape convoluted with the instrumental 
resolution.  The data shows a clear softening and dampening of this mode 
with increasing temperature until the optic peak is essentially 
unresolvable around $T_{c}$ in Fig.~\ref{optic_1} (d).  
We have also confirmed that the soft optic 
mode is unobservable above $T_{c}$ up to at least 550 K.

	The phonon frequency squared, $(\hbar \omega_{\circ})^{2}$, and the 
half-width are plotted in Fig.~\ref{optic_2}.  The square of the phonon 
frequency, $\omega_{\circ}^{2}$, is related to the dielectric constant 
$\epsilon$ via the Lyddane-Sachs-Teller (LST) relation which states that 
$(1/\epsilon)$ $\propto$ ($\hbar \omega_{\circ})^{2}$. The linear 
relationship between the square of the phonon frequency and temperature 
has been established both above and below $T_{c}$ in conventional 
ferroelectrics~\cite{Shirane67:19} and is explained in the theory of 
Cochran~\cite{Cochran69:18}.  A linear recovery of the soft optic mode has 
been observed in PMN at low temperatures.~\cite{Wakimoto02:65} These 
combined results for both PMN and PZN suggests the existence of a  
well-defined ferroelectric distortion over the length scale characterized 
by the resolution ($\sim$ 100 \AA).  
Therefore these results point to at least a 
local ferroelectric polarization in the unpoled systems.  Local or 
microdomain polarization in PZN has also been observed by Raman 
and optic studies~\cite{Lebon01:89} and it was suggested that the unpoled 
system had only orientational order and not translational order.

	As $T_{c}$ is approached from low temperatures, 
the optic mode becomes gradually more damped and 
above the critical temperature it becomes completely overdamped and is 
unobservable up to 550 K, the highest temperature studied.  This is the 
``waterfall'' region where the optic mode becomes overdamped as a result 
of the formation of polar nanoregions.  Polar nanoregions were first 
suggested by Burns and Dacol~\cite{Burns83:10} based on changes in 
the index of refraction below a high temperature denoted as the Burns 
temperature $T_{d}$.  They found the data to be well described by local 
regions of ferroelectric order in a paraelectric background. In this 
temperature range, the uncorrelated polar nanoregions with random 
polarization prevent the propagation of the optic mode.    This effect has 
been clearly observed in PMN where $T_{d}\sim$ 620 K is easily 
accessible~\cite{Gehring01:87} illustrating that the temperature 
dependence of the soft optic mode in PZN is qualitatively similar to that 
in PMN.  The waterfall effect and polar nanoregions are reviewed elsewhere 
in the context of coupled mode theory~\cite{Gehring01:63} and has been 
verified by other groups~\cite{Tomeno01:70} in this system.

	The recovery of the soft optic mode below $T_{c}$ is found to be 
quantitatively analogous to the recovery observed in PMN.  
In Fig.~\ref{optic_3} we plot $(\hbar \omega_{\circ})^{2}$ as a function 
of the reduced temperature $t=T/T_{c}$ for both PMN and PZN showing that 
the slope of the phonon recovery is the exactly same in both PMN and PZN.  We also reproduce the high-temperature data for the dielectric constant in PMN. 
The data for PMN are taken from Wakimoto \textit{et al.}~\cite{Wakimoto02:65} and the data for the high-temperature dielectric constant ($1/\epsilon$) is taken from Viehland \textit{et al.}~\cite{Viehland92:46} 
Since the soft optic mode energy characterizes the degree of the ferroelectric 
distortion, Fig.~\ref{optic_3} 
demonstrates that both PZN and PMN develop an analogous ferroelectric 
distortion at low temperatures.

\subsection{Acoustic Mode}

\begin{figure}[t]
\includegraphics[width=8cm] {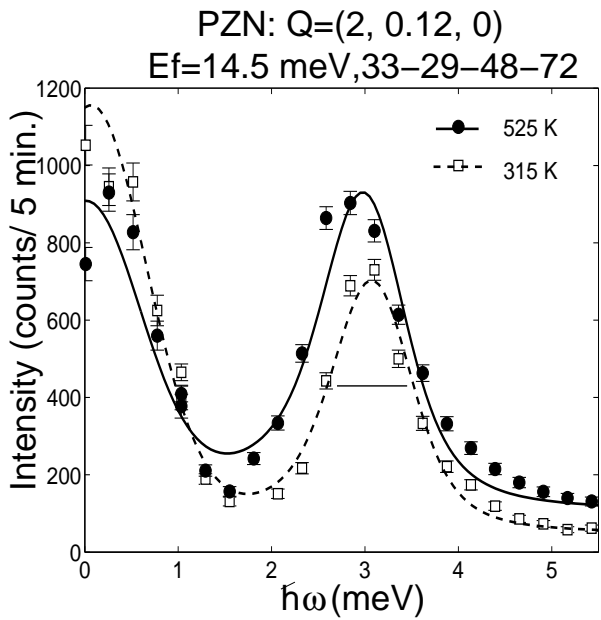}
\caption{\label{acoustic_1}Scans of the transverse acoustic (TA) mode 
above and below $T_{c}$ showing a small broadening above the critical 
temperature $\sim$ 400 K. The solid and dashed lines are the results of 
least-squares 
fits to the lineshape discussed in the text convoluted with the instrument 
resolution.  The solid bar indicates the calculated instrumental 
resolution FWHM.}
\end{figure}

\begin{figure}[t]
\includegraphics[width=8cm] {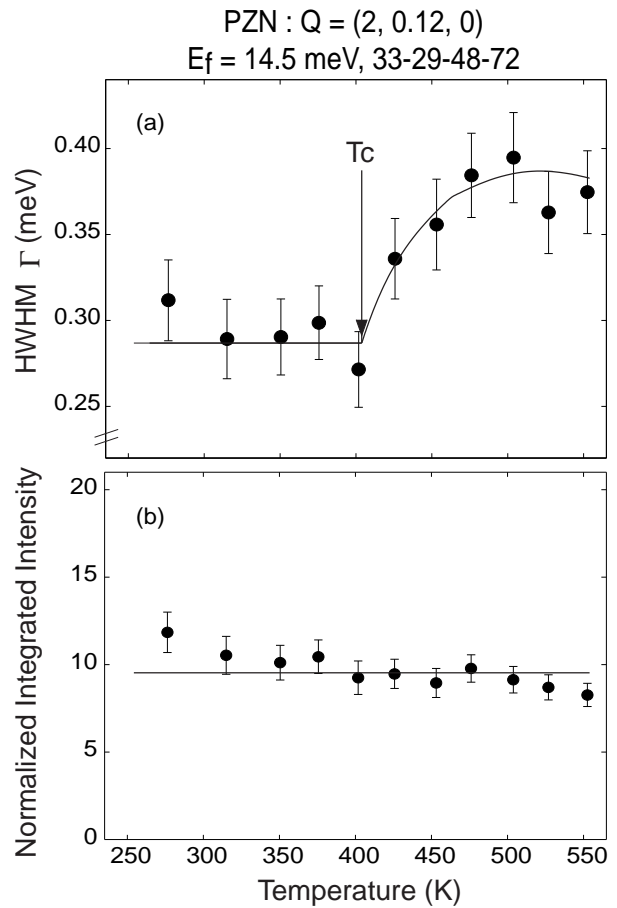}
\caption{\label{acoustic_2} Fitted parameters as a function of 
temperature.  Figure ({\it a}) 
shows the HWHM $\Gamma$ as a function of temperature.  The integrated 
intensity is plotted in figure (\textit{b}) and has an essentially flat 
temperature dependence.  The solid lines are guides to the eye.}
\end{figure}

	Studies of the transverse acoustic mode (TA) were made by conducting 
scans with fixed {\bf{Q}} at the (2, 0.12, 0) position.  Representative 
scans above and below $T_{c}$ are displayed in Fig. \ref{acoustic_1} where 
the solid and dashed curves are the results of least-squares fit to the 
lineshape previously described convoluted with the resolution.  A 
comparison of the results above and below $T_{c}$ shows a slight 
broadening at high temperatures.

	The half-width is plotted as a function of temperature in Fig. 
\ref{acoustic_2} (a) and shows a sharp anomaly at $T_{c}$ while the integrated 
intensity is essentially constant over the temperature range studied.  The 
recovery of the transverse acoustic mode for temperatures below $T_{c}$ is 
similar to that observed in PMN~\cite{Wakimoto02:65} and PT doped 
PMN~\cite{Koo02:65}. In both these cases the dampening of the acoustic mode 
appears below $T_d$ and is attributed to the formation of polar nanoregions. 
The recovery of the acoustic mode is 
associated with the development of local correlations 
between polar nanoregions resulting in local ferroelectric polarization. 
Since the acoustic modes are associated with center of mass motion they 
are less sensitive to the random field associated with the polar 
nanoregions and therefore should only be slightly broadened compared to 
the complete overdamping of the optic mode already presented.  

	Also, the recovery of both the optic and acoustic modes at $T_{c}$ suggests 
the coupling of these modes, and therefore the dynamics should be well 
described by coupled-mode theory.~\cite{Axe70:1} This was observed to be 
the case in PMN where the temperature and the zone dependent lattice 
dynamics could be well described by coupled mode theory.~\cite{Wakimoto02:xx}  
In PMN the 
appearance of diffuse scattering at $T_{d}$ was interpreted in terms 
of the condensation of the coupled soft mode.  Even though we are not able 
to probe temperatures near the Burns temperature $T_{d}$ the strong 
analogy with PMN suggests that the broadening observed in the 
acoustic mode starts at $T_{d}$ and is associated with the condensation of 
the coupled soft mode, showing another similarity 
between PZN and PMN. 

\section{Discussion and Conclusions}

	We have shown that PMN and PZN are essentially identical in terms of the temperature dependence of the diffuse scattering and the phonon scattering.  The temperature dependence of the diffuse scattering is suggestive of an onset not at the critical temperature $T_{c}$, but at the much higher Burns temperature $T_{d}$ where polar nanoregions are formed.  The temperature dependence of the phonon scattering has also been shown to be exactly analogous to that in PMN.  The recovery of the soft optic mode in PZN is identical to the observed recovery in PMN.  At temperatures above $T_{c}$, the optic mode becomes completely overdamped due to the polar nanoregions and the acoustic mode also becomes slightly damped.  Both of these results are identical to those found in the PMN system.

	Despite these similarities, the Bragg peaks in PZN show the onset of both transverse and longitudinal broadening at $T_{c}$ parallelled by a large growth in the integrated intensity.  
We have argued based on a direct comparison of our neutron results to the previous high energy x-ray results of Xu \textit{et al.} that this indicates the presence of structural inhomogeneity, probably in the bulk of the sample.  As illustrated in Fig.~\ref{figure1} PMN does not show any anomaly in the integrated intensity.  No change is also observed in the longitudinal width and only a subtle broadening is observed in the transverse width.~\cite{Wakimoto02:xx}  This suggests that the key difference between PMN and PZN is the presence of macroscopic structural disorder at low temperatures in PZN.

	A similar discrepancy in the structural properties as a function of incident x-ray energy has been observed in the doped PZN-PT system where the physical properties have been observed to vary as a function of sample thickness.  This has been directly shown by Noheda \textit{et al.} in the PZN-PT system and is illustrated by  several experiments using both high and low energy x-rays.~\cite{Noheda01:86, Noheda02:65,Noheda02:267} Our result, combined with those of previous x-ray studies conducted at various energies, provides further support to the growing evidence that structural inhomogeneity is a general feature of PZN and possibly to all relaxor systems.  This point is further illustrated by the neutron powder diffraction work by Iwase \textit{et al.} suggesting the existence of two structural phases below $T_{c}$.~\cite{Iwase99:60,Fuji00:69}   We believe that previous probes sensitive to near-surface effects, like low energy x-rays with a small penetration depth and powder diffraction, showing rhombohedral order have been interpreted as a bulk result and therefore marking a clear difference between the PMN and PZN systems, whereas in fact the bulk structural phases are quite similar.

	The recovery of the soft optic mode at low temperatures in both PMN and PZN points to the presence of ferroelectric polarization on a length scale of at least $\sim$ 100 \AA\ (defined by the resolution) in both systems despite the clear absence of any rhombohedral order in the bulk phase (phase X).  This suggests that the ferroelectric order parameter (the atomic shift or the polarization) is decoupled from the unit cell shape.  This same effect was recently observed in strained SrTiO$_{3}$ films.~\cite{He03:317}  In this system strong superlattice peaks characteristic of a structural transition were observed without any change in the c-axis lattice parameter.   This was interpreted as the internal degrees of freedom (in that case the TiO$_{6}$ rotations) becoming uncoupled from the overall lattice shape.  In PZN and PMN a ferroelectric polarization is indicated by the recovery of the soft optic mode without the presence of a rhombohedral distortion.  This fact suggests that the atomic shift is decoupled from the unit cell shape in phase X.  This picture is similar to that of Lebon \textit{et al.}~\cite{Lebon01:89} who have suggested that local orientational order (microdomains) characterizes the low temperature ground state (phase-X).  We note though that phase-X is structurally well ordered as characterized by the resolution limited Bragg peaks observed by Xu \textit{et al.} using high energy x-rays.

	Therefore, the characteristic feature of relaxors PZN and PMN is possibly a weak coupling between the ferroelectric order parameter and the structure as seen in the unit cell shape.  In the case of conventional ferroelectric systems like BaTiO$_{3}$ the coupling is strong resulting in a well defined structural transition with long-range translational and orientational order.  In the case of PMN the coupling is weaker resulting in the development of local ferroelectric polarization without any change in the unit cell shape.  In PZN, the coupling is still weaker; thus the bulk phase shows no change in unit cell.  However the coupling is slightly larger than that of PMN, therefore resulting in a well defined change in the unit cell shape with the aid of some extra strain as found near the surface of a sample. 

	The strong similarity between PMN and PZN and the nature of the relaxor phase transition in both materials, as discussed above, point to a common model for the relaxor transition.  Hereafter we discuss a universality class which possibly describes the relaxor transition from the viewpoint of an analogy of the polarization vector to magnetic spins.  When trying to formulate a model for the relaxor transition it is important to consider that there are two key temperatures which need to be described; the Burns temperature $T_{d}$ and the lower critical temperature $T_{c}$.  Also, the model needs to explain the reason for the history dependence of measured quantities around $T_c$.  To satisfy these constraints, we propose the three-dimensional Heisenberg model with cubic anisotropy universality class in the presence of random fields.  Variants of this model have been considered previously although our particular formulation and analysis appear to be unique.~\cite{Westphal92:68,Pirc99:60,Fisch03:67}  The mapping is as follows: the Heisenberg spin corresponds to the local ferroelectric polarization, the cubic anisotropy reflects the preferential orientation of the polarization along the $\langle$111$\rangle$ axes and the isotropic random magnetic field corresponds to the randomly oriented local electric fields originating from the differing charges of the Nb$^{5+}$ and Zn$^{2+}$ (Mg$^{2+}$).  Following a suggestion by Aharony~\cite{Aharony:private}, we consider the case where the Heisenberg term $H_{Heisenberg}$ dominates over the random field $H_{RF}$ which is larger than the contribution from the cubic anisotropy $H_{Cubic}$ ($H_{Heisenberg}>H_{RF}>H_{Cubic}$).~\cite{case2}

	In this scenario there would be two distinct temperature regions.  At high temperatures the cubic anisotropy is irrelevant and therefore the system should behave like a Heisenberg model in a random field.  In such a case the excitation spectrum is characterized by Goldstone modes and therefore no long-range order is expected in the presence of random fields.~\cite{Imry75:35}  The second temperature scale appears at low temperatures where the cubic anisotropy becomes relevant and therefore the system should be similar to that of an Ising system in the presence of a random field.  In such a system we would expect long-range order to occur at equilibrium or for cooling in the presence of random fields, which typically yields a non-equilibrium state, local order with history dependent effects. 

	In the actual relaxor materials PMN and PZN, the Burns temperature $T_d$ marks the entry into the higher temperature region discussed above.  Therefore, since the order parameter has a continuous symmetry~\cite{Westphal92:68} and the excitation spectrum is characterized by gapless modes, the system does not order but forms polar nanoregions in a paraelectric background.  On the other hand, the critical temperature $T_{c}$ characterizes the lower temperature region where the the system behaves more like an Ising model in the presence of a random field.   
This explains the local ferroelectric distortion characterized by the recovery of the soft-optic mode.  Even though, in equilibrium, a 3D Ising system should have long-range order in the presence of small random fields, as found in magnetic systems, non-equilibrium effects with long time scales become dominant.  The presence of non-equilibrium effects may explain the lack of true long-range order at low temperatures and the history dependence of physical properties like the linear birefringence.  We also note that the presence of the phase-shifted polar nanoregions may also pose another energy barrier for the ordered phase at the critical temperature.

	The low-temperature properties (below T$_{c}$) of PZN and PMN are exactly analogous to 
that of Mn$_{0.5}$Zn$_{0.5}$F$_{2}$ in a magnetic field~\cite{Hill91:66, Hill97:55} which 
is well described by a three-dimensional random field Ising model.  In this system the 
bulk low temperature phase is characterized by non-equilibrium effects and therefore a 
disordered metastable state was observed.  Importantly, an ordered phase was observed 
in the near-surface region of the sample, exactly analogous to the rhombohedral order 
observed in the near-surface region of PZN.  The strong similarity between these two 
systems further affirms the model we have proposed.  Our low-temperature phonon results 
are also qualitatively consistent with the observation of well-defined spin-waves in 
Mn$_{0.5}$Zn$_{0.5}$F$_{2}$ in the presence of random fields.~\cite{Leheny03:32}  
A detailed, quantitative analysis of this model would be extremely valuable.

	We have shown that PMN and PZN are very similar relaxor systems through identical behavior in the diffuse and phonon scattering.  We do observe 
structural inhomogeneity at low temperatures in the PZN system in contrast 
to PMN.  We speculate this difference is the result of stronger coupling 
in PZN between the ferroelectric order parameter and the unit cell shape.  We suggest a consistent explanation for the relaxor transition in terms of the three-dimensional Heisenberg model with cubic anisotropy in the presence of random fields.

\begin{acknowledgements}

	We thank A. Aharony for pointing out one of the key concepts introduced in this paper.  We also are 
grateful to P. M. Gehring and G. Xu for useful discussions and M. M. Potter, 
J. J. -P. Bolduc, and L. E. McEwan for technical support. The work at the 
University of Toronto was supported by the Natural Science and Engineering 
Research Council of Canada.  We acknowledge financial support from the U. 
S. DOE under contract No. DE-AC02-98CH10886, and the Office of Naval 
Research under Grant No. N00014-99-1-0738.

\end{acknowledgements}


\end{document}